\documentclass[%
 reprint,
superscriptaddress,
nofootinbib,
 amsmath,amssymb,
 aps,
]{revtex4-1}

\usepackage{graphicx}
\usepackage{dcolumn}
\usepackage{bm}
\usepackage{hyperref}

\begin{document}

\newcommand{\infnto}{Istituto Nazionale di Fisica Nucleare (INFN), Sezione di Torino, Via P.\ Giuria 1, I-10125 Turin, Italy}
\newcommand{\ific}{Instituto de F{\'\i}sica Corpuscular  (CSIC-Universitat de Val{\`e}ncia), E-46980 Paterna, Spain}
\newcommand{\infnfe}{Istituto Nazionale di Fisica Nucleare (INFN), Sezione di Ferrara, Via G.\ Saragat 1, 44122 Ferrara, Italy}
\newcommand{\aarhus}{Department of Physics and Astronomy, Aarhus University, Ny Munkegade 120, DK-8000 Aarhus C, Denmark}
\newcommand{\kit}{Institut f\"ur Astroteilchenphysik, Karlsruhe Institute of Technology (KIT),
Hermann-von-Helmholtz-Platz 1, 76344 Eggenstein-Leopoldshafen, Germany}
\newcommand{\iitb}{Department of Physics, Indian Institute of Technology Bombay, Powai, Mumbai 400076, India}
\newcommand{\UT}{Department of Physics, University of Texas, Austin, TX, USA}
\newcommand{\OKC}{OsKar Klein Centre for Cosmological Physics, Stockholm University, Stockholm, Sweden}
\newcommand{\unife}{Dipartimento di Fisica e Scienze della Terra, Universit\`a di Ferrara, Via G.\ Saragat 1, 44122 Ferrara, Italy}

\title{Neutrino mass and mass ordering: No conclusive evidence for normal ordering}

\author{Stefano Gariazzo}
\affiliation{\infnto}

\author{Martina Gerbino}
\affiliation{\infnfe}

\author{Thejs Brinckmann}
\affiliation{\infnfe}
\affiliation{\unife}

\author{Massimiliano Lattanzi}
\affiliation{\infnfe}

\author{Olga Mena}
\affiliation{\ific}

\author{Thomas Schwetz}%
\affiliation{\kit}

\author{Shouvik Roy Choudhury}%
\affiliation{\iitb}

\author{Katherine Freese}%
\affiliation{\UT}
\affiliation{\OKC}

\author{Steen Hannestad}
\affiliation{\aarhus}

\author{Christoph A. Ternes}%
\affiliation{\infnto}

\author{Mariam T{\'o}rtola}%
\affiliation{\ific}

\newcommand{\mnu}{\ensuremath{\Sigma m_\nu}}
\newcommand{\mnuev}{\ensuremath{\Sigma m_{\rm\nu, eV}}}
\newcommand{\dmsol}{\ensuremath{\Delta m^2_{21}}}
\newcommand{\dmatm}{\ensuremath{|\Delta m^2_{31}|}}
\newcommand{\DKL}{D_\mathrm{KL}}

\begin{abstract}
The extraction of the neutrino mass ordering is one of the major challenges in particle physics and cosmology, not only for its implications for a fundamental theory of mass generation in nature, but also for its decisive role in the scale of future neutrinoless double beta decay experimental searches.
It has been recently claimed that current oscillation, beta decay and cosmological limits on the different observables describing the neutrino mass parameter space provide robust \emph{decisive} Bayesian evidence in favor of the normal ordering of the neutrino mass spectrum~\cite{Jimenez:2022dkn}.
We further investigate these strong claims using a rich and wide phenomenology, with different sampling techniques of the neutrino parameter space.
Contrary to the findings of Jimenez \emph{et al}~\cite{Jimenez:2022dkn}, no decisive evidence for the normal mass ordering is found.
Neutrino mass ordering analyses must rely on priors and parameterizations that are ordering-agnostic: robust results should be regarded as those in which the preference for the normal neutrino mass ordering is driven exclusively by the data, while we find a difference of up to a factor of 33 in the Bayes factors among the different priors and parameterizations exploited here.
An ordering-agnostic prior would be represented by the case of
parameterizations sampling over the two mass splittings and a mass scale,
or those
sampling over the individual neutrino masses \emph{via normal prior distributions only}.
In this regard, we show that the current significance in favor of the normal mass ordering should be taken as $2.7\sigma$ (i.e.\ \emph{moderate} evidence), mostly driven by
neutrino oscillation data.
\end{abstract}

\maketitle

\section{Introduction}
Understanding the origin of elementary particle masses is a key question in particle physics.
Neutrino masses are known to be below the eV scale and therefore many orders of magnitudes smaller than those of charged fermions.
A precise knowledge of the leptonic mixing sector is a mandatory step to build a complete theoretical framework that explains the origin of masses in nature.
A crucial and essential ingredient is the extraction of the neutrino mass ordering: \emph{normal} versus \emph{inverted}.
These two possible options are due to the fact that the sign of the atmospheric mass splitting ($|\Delta m^2_{31}| \approx 2.5\cdot 10^{-3}$~eV$^2$~\cite{deSalas:2020pgw,Esteban:2020cvm,Capozzi:2021fjo}) remains unknown~\cite{DeSalas:2018rby}.

Neutrino oscillations are sensitive to the mass ordering due to the MSW matter effect \cite{Wolfenstein:1977ue,Mikheev:1986gs}, an interference effect between the two mass-squared splittings~\cite{Petcov:2001sy}, or by the comparison of $\nu_e$ and $\nu_\mu$ disappearance measurements \cite{Nunokawa:2005nx,Blennow:2012gj}, and the determination of the sign of $\Delta m^2_{31}$ is one of the major physics goals of the next generation of oscillation experiments \cite{DUNE:2020jqi,Hyper-Kamiokande:2018ofw,JUNO:2015zny,IceCube-Gen2:2019fet}, see also \cite{Blennow:2013oma,DeSalas:2018rby}.
Cosmological measurements also provide a suitable framework to test the neutrino mass ordering, see, e.g.~\cite{Lesgourgues:2013sjj,Hannestad:2016fog,Vagnozzi:2017ovm}.
In the minimal mass scenario (vanishingly small mass of the lightest neutrino state), measurements of the neutrino mass splittings by flavour oscillation experiments imply a lower limit to the sum of the neutrino masses.
In the normal ordering, the sum of the neutrino masses obeys $\sum m_\nu \gtrsim 0.06$~eV, while, in the inverted ordering, $\sum m_\nu \gtrsim 0.10 $~eV.
By bounding the sum of neutrino masses from above, cosmology could exclude the inverted mass ordering, even though the mass ordering is not directly measurable with current (or near future) data.

When only oscillation measurements are considered, global fit analyses provide a $2-2.7\sigma$~\cite{deSalas:2020pgw,Esteban:2020cvm,Capozzi:2021fjo} preference for the normal ordering.
This significance is only mildly increased (up to $0.7\sigma$) when the most aggressive cosmological data sets are added to the oscillation results~\cite{deSalas:2020pgw,Capozzi:2021fjo}.

Extracting the neutrino mass ordering is of high relevance: apart from the pure theoretical perspective, the mass ordering rules the scale of future neutrinoless double beta decay searches.
As a consequence, a race to establish the mass ordering from current observations seems to have started in the literature, with some controversy in the results.
Namely, a \emph{strong evidence} for the normal neutrino mass
ordering was claimed in~\cite{Simpson:2017qvj}, where Bayesian odds of 42:1 in favour of the normal ordering were reported.
These results were discussed in \cite{Schwetz:2017fey}, where the authors explained them as a consequence of the particular choice of a flat logarithmic prior for the three neutrino masses.
A reassessment of this criticism was fully developed in \cite{Gariazzo:2018pei}:
The evidence for normal ordering was found to be strong only when the scan was performed over the three neutrino masses with logarithmic priors.
For every other possible parameterization and/or prior choice explored in the paper, the preference was merely weak, and driven by neutrino oscillation data.
The role of priors in the inference of neutrino masses and ordering was also recently discussed in \cite{Hergt:2021qlh}.

More recently, the authors of Ref.~\cite{Jimenez:2022dkn} have claimed \emph{decisive} Bayesian evidence for the normal neutrino mass ordering.
We scrutinize here these very strong and, apparently, robust findings by means of different sampling parameters and choices of prior distributions to describe the neutrino parameter space.
Within the five different sampling choices considered here,
we observe a significant variation of the preference in favor of NO,
which ranges from weak/moderate to strong and is dominated by neutrino oscillation results,
while stronger limits on the sum of the neutrino masses make the preference only mildly more significant.

The structure of this manuscript is as follows.
In Sec.~\ref{sec:analysis}, we describe the analysis methodology and present the statistical approach, the parameterizations and the data we used.
We show our results in Sec.~\ref{sec:results}, followed by a discussion on ordering-agnostic priors and parameterizations, i.e.\ not yielding an \emph{a priori} preference for a given mass ordering scheme, in Sec.~\ref{sec:agnostic}.
We conclude in Sec.~\ref{sec:conclusions}.

\section{Analysis setup}
\label{sec:analysis}
In this section, we present our numerical implementation and the data constraints we used,
with the purpose of discussing the impact of subjective parameterization and prior choices on the mass ordering preference.

\subsection{Statistical method}

The comparison between normal ordering (NO) and inverted ordering (IO) is performed by means of the Bayes factors
$B=Z_{\rm NO}/Z_{\rm IO}$, or, equivalently, $\ln B=\ln Z_{\rm NO}-\ln Z_{\rm IO}$,
most commonly used as a reference in deciding the level of significance.
Here $Z_{\rm NO, IO}$ denotes the Bayesian evidence for NO or IO, respectively.
We define the significance of our results by means of the probability in favor of NO
instead of using the Jeffreys' scale or an equivalent one.
We assume a prior probability of 50\% for both NO and IO.
Under this assumption, after data are considered and the Bayes factor is computed,
the probability in favor of NO becomes $p=B/(1+B)$,
which can be converted to a significance in terms of a number of sigma for a Gaussian variable
using the inverse error function, see e.g.~\cite{DeSalas:2018rby}~\footnote{
Notice that the conversion performed in this way generally provides smaller significances
as compared to the $\chi^2$ distribution with 1 degree of freedom, commonly used
in a frequentist context (see \cite{Blennow:2013oma} for a discussion).
For instance, a $\Delta\chi^2=9$ in favor of NO would correspond to a $3\sigma$ significance
in a frequentist sense,
but it corresponds to $B=e^{9/2}$
(provided that the likelihood satisfies $-\ln\mathcal{L}=\Delta\chi^2/2$),
$p=98.9$\% and consequently a significance of $2.5\sigma$ with our Bayesian definition.}.

We use PolyChord~\cite{Handley:2015aa} to compute the Bayesian evidences
of different parameterizations (see later), considering some simplified constraints
that approximate real neutrino oscillation, terrestrial and cosmological measurements of neutrino masses and mass differences.
For each configuration, we repeat the calculation several times in order to reduce
the impact of statistical fluctuations in the selection of the live points,
which affects mostly parameterizations with three free neutrino masses.
The results we show in the following are obtained from the mean and standard deviation
of the natural logarithm of the Bayesian evidences we get from several runs,
for which we use from 500 to 1500 live points.
Notice that in some cases the variance is so small that error bars are not visible in the plots.

\subsection{Parameterizations}
\label{subs:param}

In order to sample the parameter space for neutrino masses,
we have several possibilities.
In this work, we have adopted five different parameterizations that are fairly representative of the full range of possibilities~\footnote{We have not considered parameterizations and prior choices that are rooted in specific theories of neutrino mass generation such as that proposed in~\cite{Long:2017dru}. Their inclusion would not change the conclusions of this work.}:
\begin{itemize}
\item (\textbf{A}) Following the PolyChord capability of performing Bayesian evidence calculations in case of unbounded parameters,
the first parameterization we consider is described by a Gaussian prior on the logarithm
of the three neutrino mass eigenstates $m_A$, $m_B$, $m_C$ (a lognormal prior).
The Gaussian prior is described by two hyperparameters of the analysis,
$\mu$ and $\sigma$,
which are marginalized over.
The mean and standard deviation of the Gaussian are $\ln\mu$ and $\sigma$, respectively~%
\footnote{In other words: $\ln (\mu/\mathrm{eV}) = \mathrm{E}[\ln (m/\mathrm{eV})]$ and $\sigma^2 = \mathrm{Var}[\ln ( m/\mathrm{eV})]$.}
(notice that this is the standard deviation of the logarithm of the mass scale).
While sampling, we consider a uniform prior on the logarithm of $\mu$ and $\sigma$, with bounds
$5\cdot10^{-4}<\mu/\text{eV}<0.3$ and $5\cdot10^{-3}<\sigma<20$, respectively.
Once the sampling of the three masses is performed, we order them and assign
the values to the mass eigenstates $m_1$, $m_2$, $m_3$ according to the considered mass ordering. Note that this parameterization is based on Refs.~\cite{Jimenez:2022dkn,Simpson:2017qvj}.
\item In order to maintain a closed parameter space, alternatively,
we sample the neutrino masses within a given range ($[10^{-6}, 10]$ in this case),
either uniformly on the masses (case \textbf{B}) or their logarithms (case \textbf{C}).
On top of the uniform sampling, we apply a Gaussian constraint
with the same mean and standard deviation we discussed as hyperparameters in the previous case.
The two parameters have the same logarithmic prior as before and are marginalized over.
Finally, we sort and assign the masses given the considered mass ordering as previously described.
We verified that sampling the masses independently or imposing that they are sorted
from smallest to largest at sampling time has no impact on the results.
\item (\textbf{D}) Following \cite{Gariazzo:2018pei},
another parameterization we consider involves the lightest neutrino mass and the two
mass splittings \dmsol\ and \dmatm.
We consider the same uniform prior distribution on the logarithm of the three parameters,
with bounds $[10^{-6}, 10]$.
We comment later in the text on the impact of using a uniform prior on the parameters.
\item (\textbf{E}) Finally, we describe the neutrino masses by means of their sum \mnu\ and the two
mass splittings \dmsol\ and \dmatm\ \cite{Heavens:2018adv}.
The three of them are sampled from a uniform distribution,
with bounds $[10^{-6}, 10]$.
Given the values of the two mass splittings, however,
part of the parameter space available to \mnu\ corresponds
to unphysical masses and is therefore removed.
\end{itemize}

We checked that altering the above-mentioned prior boundaries to $[10^{-5}, 1]$
has almost negligible effects on the Bayes factors in the last two cases,
while it may alter significantly
the results obtained when considering three neutrino masses as free parameters.
A detailed comparison of such effects, however,
is beyond the scope of this analysis,
since the first three parameterizations still provide a much larger preference than the last two.

\subsection{Data constraints}

We consider results from recent global fits to neutrino oscillation data~\cite{deSalas:2020pgw,Capozzi:2021fjo,Esteban:2020cvm},
from $\beta$-decay constraints on the effective electron neutrino mass by KATRIN~\cite{KATRIN:2021uub}
and from cosmological observations~\cite{Planck:2018vyg,DiValentino:2021hoh,Palanque-Delabrouille:2019iyz}.
The scope of this analysis is to demonstrate
the effect of priors on the final results.
A simplified setup that runs quickly is suitable
for the purpose of performing several tests and commenting on the robustness of the conclusions
based on different parameterizations, priors or sampling choices.
Therefore, we adopt an approximate formulation of the experimental likelihoods,
implemented as follows:

\begin{itemize}
\item Neutrino oscillation constraints are accounted for with a Gaussian likelihood
on the solar mass difference, $\Delta m^2_{21}\approx(7.5\pm0.2)\cdot10^{-5}$~eV$^2$,
and on the absolute value of the atmospheric mass difference
$|\Delta m^2_{31}|\approx(2.50\pm0.03)\cdot10^{-3}$~eV$^2$.
Notice that we do impose the same Gaussian constraints on both mass orderings.
\item Terrestrial neutrino mass measurements are accounted for with a Gaussian likelihood, whose parameters are set according to
KATRIN results \cite{KATRIN:2021uub}:
\begin{equation}
\ln\mathcal{L}_{\rm KATRIN}
=
-\frac{1}{2}
\left(\frac{0.1-(\mnuev/3)^2}{0.3}\right)^2
\,,
\end{equation}
where $\mnuev$ is the sum of neutrino masses in eV~\footnote{In doing so, we are approximating the effective electron neutrino mass with a third of the mass sum. This approximation is perfectly justified in the range of masses currently probed by KATRIN.}.
\item Concerning cosmology, we consider two different limits:
$\mnu<0.12$~eV~\cite{Planck:2018vyg}, which is the official 95\% C.L. bound from the Planck 2018 legacy release, and
$\mnu<0.09$~eV~\cite{Palanque-Delabrouille:2019iyz,DiValentino:2021hoh}, which corresponds to the most aggressive 95\% C.L.\ bound to-date.
In both cases, we approximate cosmological information with a Gaussian likelihood centered on zero
and with the corresponding standard deviation of 0.06~eV or 0.045~eV, respectively.
Further details on the role of cosmological bounds in our analysis are reported in Sec.~\ref{sub:cosmo}.
\end{itemize}

To summarize, we consider four different data combinations:
oscillations alone,
terrestrial experiments only (oscillations+KATRIN),
and two cases with oscillation data combined with cosmological limit on the sum of the neutrino masses,
one more aggressive ($\mnu<0.09$~eV) and one slightly more conservative ($\mnu<0.12$~eV).

\section{Results on the neutrino mass ordering}
\label{sec:results}

In this section, we report our results.
We discuss the preference in favour of normal ordering quantified in terms of the Bayes factor in Sec.~\ref{sub:bayes_factor}.
In Sec.~\ref{sub:KL}, we comment on the impact of a different parameterization choice by computing the relative entropy between prior and posterior distributions, i.e., the Kullback-Leibler divergence.
We briefly comment on the role of cosmological bounds in Sec.~\ref{sub:cosmo}.

\subsection{Bayes factor}\label{sub:bayes_factor}

\begin{figure*}
    \centering
    \includegraphics[width=\textwidth]{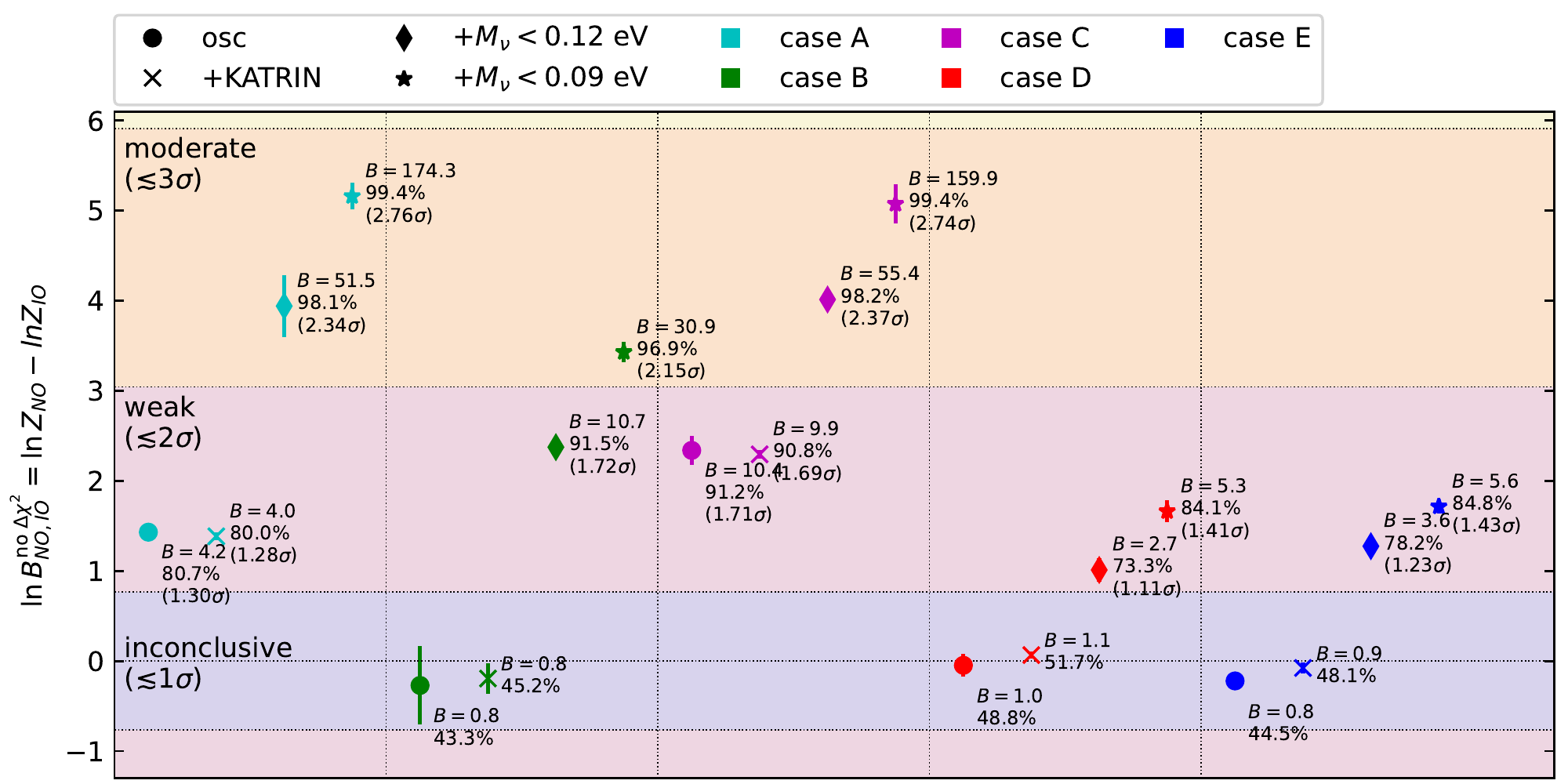}
    \caption{Bayes factors $B$ in favor of NO:
    probability of NO being the true mass ordering and corresponding number of $\sigma$
    (not indicated if smaller than $1\sigma$),
    for different parameterizations and constraining data, as described in the text.
    Colored bands represent the Bayes factor ranges that correspond to statistical significance
    of $1\sigma$, $2\sigma$ and $3\sigma$ Gaussian probabilities.
    Note that the $\Delta\chi^2$ value from oscillation experiments has not been included in this figure.}
    \label{fig:Bayes_factors}
\end{figure*}

Figure~\ref{fig:Bayes_factors} summarizes our results concerning the comparison between NO and IO,
when the neutrino oscillation preference for NO is not taken into account.
We show the Bayes factors plus the corresponding probability and significance in favor of NO
for the five different parameterizations described in section~\ref{subs:param}.

As we can see, considering only oscillation data (circles) or adding the KATRIN constraints (crosses)
on the effective
electron neutrino mass gives the same results (apart from small numerical fluctuations):
the limit on the neutrino masses from KATRIN is broad enough not to alter the Bayes factor
in favor of NO.
The significance of the preference in favor of NO, however, changes significantly when we consider
different parameterizations.
Given that the oscillation $\Delta\chi^2$ between NO and IO is not taken into account
and that the constraints on the mass splittings are the same,
we expect that there is no preference in favor of NO when only oscillation data are considered,
and indeed this is what we obtain in cases \textbf{B}, \textbf{D} and \textbf{E}.
On the contrary, cases \textbf{A} and \textbf{C} provide a preference in favor of NO,
which therefore is necessarily
a consequence of the specific choice of parameters and their priors,
because the data are the same.

When more constraining observations
in the form of upper limits on the sum of neutrino masses,
$\mnu<0.12$~eV (diamonds) or $\mnu<0.09$~eV (stars), are taken into account,
the preference in favor of NO increases.
Cases \textbf{D} and \textbf{E} are almost equivalent, providing a Bayes factor just above 5,
which corresponds to approximately 85\% probability ($1.4\sigma$) in favor of NO.
Concerning the other cases, the preference can be significantly higher,
reaching a Bayes factor of approximately 175, for a nearly $2.8\sigma$ preference for NO in case \textbf{A},
i.e.\ twice the significance of cases \textbf{D} or \textbf{E}.
Notice that even when considering the three neutrino masses as free parameters (cases \textbf{A}, \textbf{B}, \textbf{C}),
the Bayes factor in favor of NO varies up to a factor 6 depending on the selected sampling method.
We can also note that when switching from terrestrial measurements only
to the terrestrial plus cosmological constraints,
the Bayes factor increases differently in the various cases:
from a factor around 40 for cases \textbf{A}, \textbf{B} to a much smaller 5 in case \textbf{D}.
Varying the parameterization or the prior, therefore, does not simply introduce a global normalization
in the Bayes factors, but rather it exacerbates the preferences in favor of NO when
the constraints on the absolute mass scale become tighter.

Notice that our results are in good agreement with those obtained in \cite{Jimenez:2022dkn}
for case \textbf{A} (their SJPV prior),
with a slightly increased significance when including cosmological constraints,
while in case \textbf{E} (their HS prior) we obtain slightly less significant results.
The minimum preference for NO when considering the strongest cosmological constraints, however,
is obtained within our case \textbf{D},
for which the preference for NO is only driven by data:
\emph{this should be considered as the most robust result.}
The effect of parameterization and prior choices, therefore,
is as large as a factor of 33 in the Bayes factor, solely due to subjective choices,
when changing from case \textbf{A} to case \textbf{D}.

\begin{figure*}
    \centering
    \includegraphics[width=\textwidth]{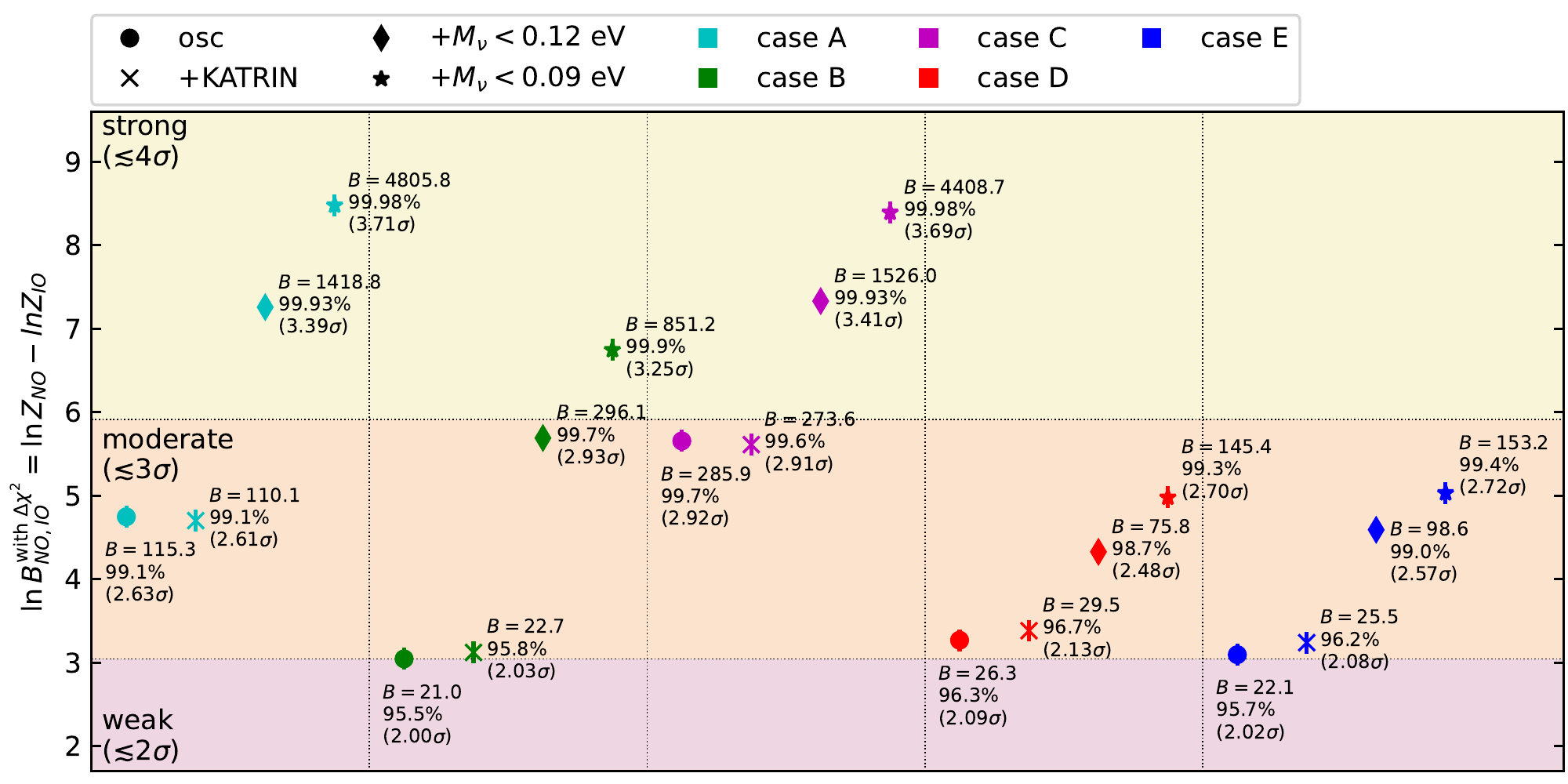}
    \caption{Same as figure~\ref{fig:Bayes_factors},
    but taking into account the additional preference in favor of NO arising from neutrino oscillations; we use the average $\Delta\chi^2$ obtained in the global analyses~\cite{Esteban:2020cvm,deSalas:2020pgw,Capozzi:2021fjo}.
    Colored bands represent the Bayes factor ranges that correspond to statistical significance
    of $2\sigma$ to $4\sigma$ Gaussian probabilities.
    }
    \label{fig:Bayes_factors_dchi2}
\end{figure*}

When including the $\Delta\chi^2$ from oscillation experiments in favor of NO~%
\footnote{This is achieved by multiplying the Bayes factor by $e^{\Delta\chi^2/2}$,
or adding $\Delta\chi^2/2$ to its logarithm.},
which is currently between 6.5 \cite{deSalas:2020pgw,Capozzi:2021fjo} and 7 \cite{Esteban:2020cvm}
with the full datasets, we obtain the preferences summarized in Fig.~\ref{fig:Bayes_factors_dchi2}.
As can be seen,
the preference in favor of NO ranges between $2.7\sigma$ (case \textbf{D}) and $3.7\sigma$ (case \textbf{A})
when the most stringent cosmological limit is considered.
The former case, in particular, is not considered as decisive in particle physics.
Notice that the oscillation $\Delta\chi^2$ alone gives a Bayes factor
of $e^{3.2}$ to $e^{3.5}$ in favor of NO,
which corresponds to a significance of $2.1$ to $2.2\sigma$.
The increase due to cosmological constraints, when a conservative parameterization is considered,
is therefore a mere $0.6\sigma$.

We conclude this section with the following note: in case \textbf{D}, we verified that a uniform prior on the parameters gives the same results as those presented in this section and obtained with a uniform prior on the logarithm of the parameters.

\subsection{Relative entropy}
\label{sub:KL}

We also assess the impact of different parameterizations by computing the relative entropy between the prior and the posterior.
The relative entropy, or Kullback-Leibler (KL) divergence, $\DKL(P||Q)$, between two probability distributions
$P$ and $Q$ of a continuous random variable $\mathbf{x}$ is given by:
\begin{equation}
\DKL(P||Q) = \int P(\mathbf{x})\ln\frac{P(\mathbf{x})}{Q(\mathbf{x})} d\mathbf{x} \, .
\end{equation}
The KL divergence is a measure of the ``distance''~\footnote{Note that the relative entropy is not strictly a distance, as it is not symmetric for exchange of $P$ and $Q$, nor does it satisfy the triangle inequality.} between the two distributions.
In the context of Bayesian inference, $\DKL(P||Q)$ measures how much information is gained when prior beliefs, encoded in the prior $Q$, are updated, after observations, to bear the posterior $P$.

\begin{figure*}
    \centering
    \includegraphics[width=\textwidth]{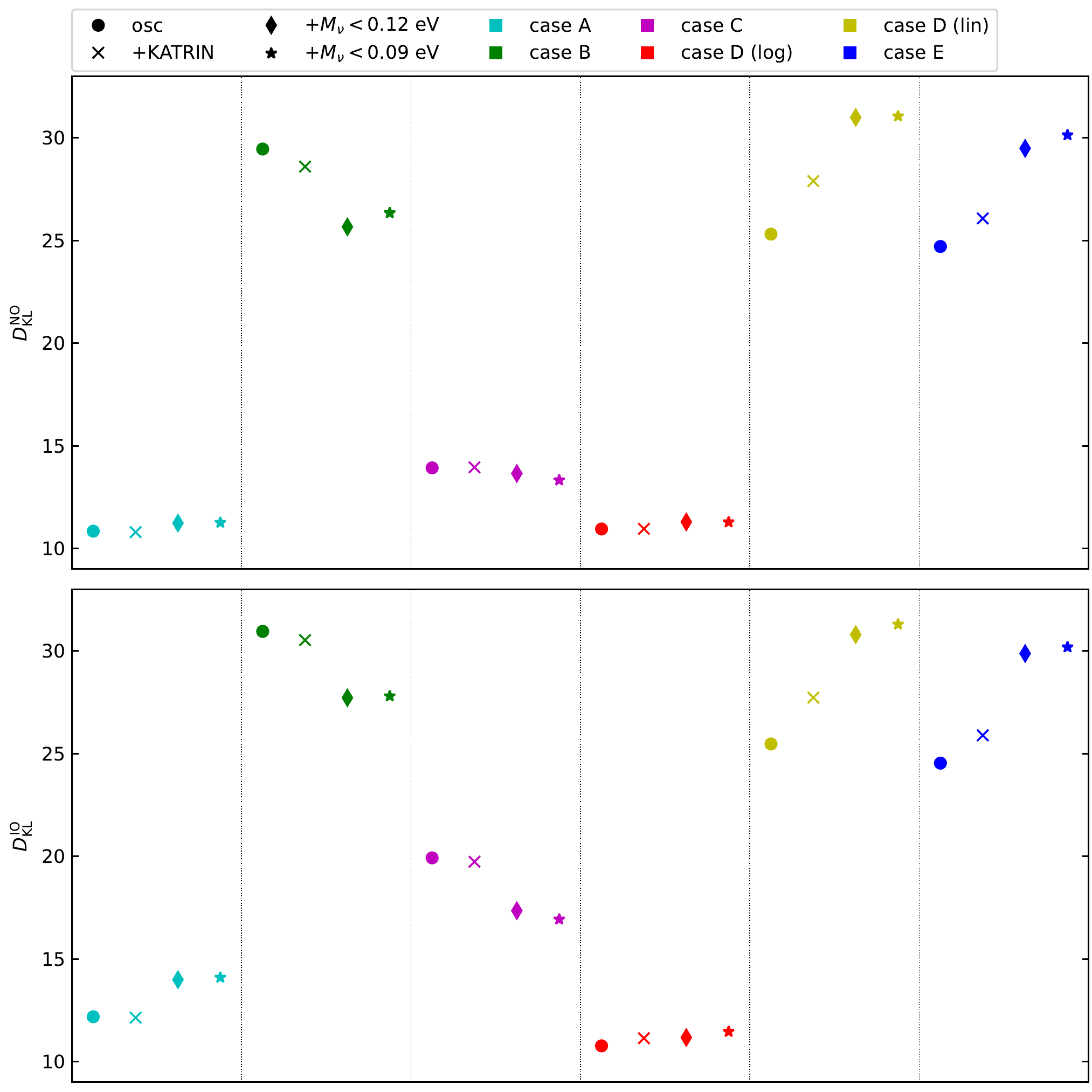}
    \caption{Kullback-Leibler divergence $\DKL$ from the prior to the posterior of the mass parameters, for NO (top panel) and IO (bottom panel), for different parameterizations and constraining data, as described in the text. Larger values of $\DKL$ signal a greater importance of the data and therefore less importance of the prior.
    }
    \label{fig:KLdiv}
\end{figure*}

We have computed, using Polychord, the KL divergence between the full (i.e., multidimensional) parameter prior and posterior for all the parameterization and dataset combination considered so far.
The results are shown in Fig.~\ref{fig:KLdiv}.
A larger value of $\DKL$ corresponds to a greater information gain.
If we compare the obtained values, from the same data, for different parameterizations (i.e., same symbol but different colors), we can gauge the impact of the parameterization itself, and of the prior induced by this choice, on the information gain~%
\footnote{In fact, the reference prior proposed in Ref.~\cite{Bernardo:1979} as an ``objective'' prior, is built from the property of maximizing, on average, the information gain.}.
In particular, less (more) informative priors lie in the top (bottom) part of the plot.
It is evident how the parameterizations in which the mass parameters are sampled linearly
(cases \textbf{B}, \textbf{D} linear and \textbf{E}) yield the largest information gain.
On the contrary, when the mass parameters are sampled logarithmically, the information gain is smaller, signaling a larger weight of the prior in the final constraints.
This also confirms the intuition that the ``uninformative'' prior is the one in which the parameters that are more directly measured by the data are sampled uniformly \cite{Bernardo:1979,Heavens:2018adv}.
The findings reported in Fig.~\ref{fig:KLdiv} also mostly confirm, from a different perspective, the conclusions of Figs.~\ref{fig:Bayes_factors} and~\ref{fig:Bayes_factors_dchi2}: for the same data, logarithmic priors tend to yield stronger preference for NO, i.e., they are more informative.
The only exception is case \textbf{D}: while for computing of the odds in favour of NO there was no impact on the results when sampling linearly or logarithmically on the mass of the lightest eigenstate and on the two mass splittings, this makes instead a difference for the KL divergence.
This is not necessarily a contradiction, as the Bayes factors shown in Figs.~\ref{fig:Bayes_factors} and~\ref{fig:Bayes_factors_dchi2} on one side, and the KL divergence shown in Fig.~\ref{fig:KLdiv} on the other, are different metrics of the information gain.
The Bayes factor is measuring how much we learn \emph{about the ordering, with respect to $1:1$ prior odds}\footnote{In the case of a random process with two possible outcomes, it is in fact straightforward to express the KL divergence between the posterior and the $1:1$ prior in terms of the Bayes factor.}.
The KL divergence shown in the figure is instead assessing
the relative information \emph{on the mass parameters, between the posterior and the prior induced by a given parametrization}.
While we expect these quantities to be related, it is not impossible that they yield different results.

\subsection{Impact of cosmological bounds}
\label{sub:cosmo}

The cosmological bounds we adopt in this work assume a minimally extended $\Lambda$CDM plus $\mnu$ model, which results in some of the most stringent constraints available.
However, cosmology is naturally model-dependent and in order to remain agnostic about the true properties of the Universe, it is quite reasonable to consider extended models and how that might affect neutrino mass sum bounds from cosmology.
This discussion is relevant because a relaxed cosmological bound on $\mnu$ would favour the non-hierarchical region of parameter space, thus clearly decreasing the odds in favour of NO.

In general, due to the effect of parameter degeneracy, relaxed bounds on $\mnu$ are obtained when considering extended models, especially those that allow for more freedom in the late-time evolution of the Universe, i.e., spatial curvature, dark energy, modified gravity (although physically motivated restrictions to the parameter space can have the opposite effect~\cite{Vagnozzi:2018jhn,RoyChoudhury:2018vnm}).
However, the relaxation never exceeds a factor of a few, keeping $\mnu$ well below the eV scale, see, e.g.,~\cite{Gariazzo:2018meg,RoyChoudhury:2019hls,DiValentino:2019dzu,Ballardini:2020iws} for recent studies.

It is also worth noting that constraints on the sum of neutrino masses from Planck might be impacted by the so-called lensing anomaly, i.e.\ the fact that the amount of peak smearing observed by the Planck satellite exceeds the theoretical expectation due to standard gravity~\cite{Planck:2018vyg}.
Thus, Planck data will tend to favor models that increase peak smearing, such as very low mass neutrinos.
A simple strategy to factor in this effect is to artificially rescale the lensing amplitude used for lensing the primary CMB spectra through the so-called $A_\mathrm{lens}$ parameter~\cite{Calabrese:2008rt}, and marginalize over it, although more refined mitigation strategies have been recently proposed~\cite{Sgier:2021bzf}.
As a result, the neutrino mass bounds can be relaxed by a factor of a few, see, e.g.,~\cite{RoyChoudhury:2019hls,DiValentino:2019dzu,Sgier:2021bzf}.

The scope of this paper is to discuss the robustness of the preference for NO, which is stronger for lower $\mnu$ limits.
Thus, in our analysis, we only consider cosmological constraints obtained in the framework of the minimal extension of the $\Lambda$CDM model with neutrino masses.
Considering weaker bounds from extended models would lessen the impact of cosmological data on the preference for NO, and bring the results closer to those obtained from oscillation data alone.

\section{On the ordering-agnostic prior}
\label{sec:agnostic}

An interesting point to make is whether a given parameterization and prior choice are agnostic (i.e., it shows no \emph{a priori} preference for one particular case/model) with respect to the mass ordering, see e.g., Refs.~\cite{Mahony:2019fyb,Gerbino:2016ehw}.
The 5 parameterizations adopted in this work and detailed in Sec.~\ref{subs:param} can be grouped in two broad classes:
i) sampling over the three individual neutrino masses;
ii) sampling over a mass-scale parameter (either the lightest mass or the mass sum) and two mass splittings.
In order to let the data drive the preference for the mass ordering, we would like that the prior choice adopted in either i) or ii) assigns equal weight to the ordering before data are taken into account.

In this subsection, we check whether this is the case by drawing a set of $N$ samples for each of the two classes defined above.
In the case of i), a given sample is assigned to represent normal ordering if $L^2-M^2>M^2-S^2$, where $L,\,M,\,S$ are the individual masses of the largest, medium and smallest eigenstate respectively.
In the case of ii), normal ordering is obtained if $\Delta m_a<\Delta m_b$, where $\Delta m_{a,b}$ are the two mass splittings.
In both i) and ii), we expect the prior choice to be ordering-agnostic if normal ordering is assigned on average to 50\% of the samples (with of course the remaining 50\% assigned to inverted ordering).

\begin{figure*}
    \centering
    \includegraphics[width=\textwidth]{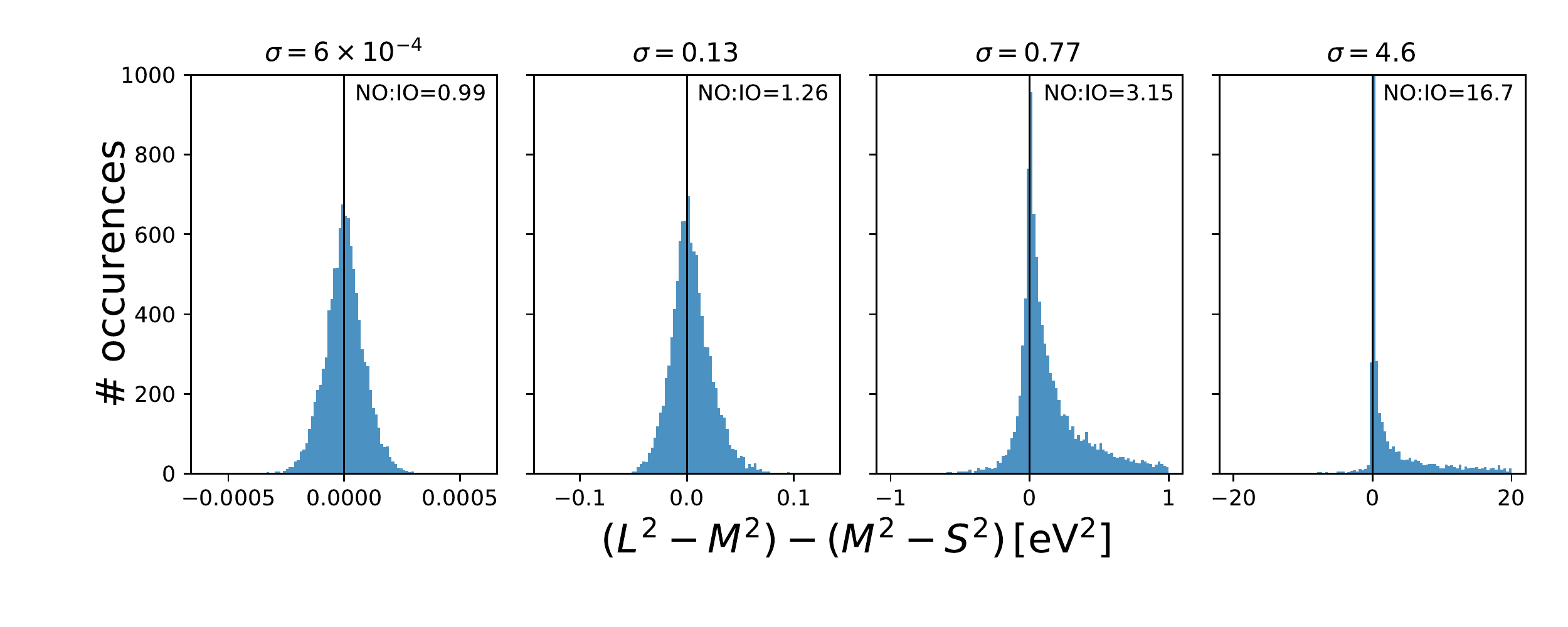}
    \caption{Empirical probability distribution of the variable $(L^2-M^2)-(M^2-S^2)$, where $L,\,M,\,S$ are the individual masses of the largest, medium and smallest eigenstates, respectively. The variable can be thought of as a proxy for the preference for normal ordering with respect to inverted ordering, with $(L^2-M^2)-(M^2-S^2)>0$ representing normal ordering. The preference is quantified as the ratio between the number of occurences corresponding to NO and IO, and reported in the top-right corner of each panel for that specific instance (an assessment of the error is quoted in the main text). The individual masses are drawn from a log-normal distribution with fixed value of $\ln \mu=-1.3$ and the values of $\sigma$ reported on top of each panel. In presence of an ordering-agnostic prior, we expect the distribution of $(L^2-M^2)-(M^2-S^2)$ to be symmetric about the origin, and $\mathrm{NO:IO=1}$.
    }
    \label{fig:hist}
\end{figure*}

In Fig.~\ref{fig:hist}, we report the empirical distribution of the variable $(L^2-M^2)-(M^2-S^2)$ when $L,\,M,\,S$ are sampled from the same log-normal prior distribution (case \textbf{A} in Sec.~\ref{subs:param}).
The four panels in Fig.~\ref{fig:hist} correspond to log-normal distributions with fixed $\ln \mu=-1.3$ and varying $\sigma$, increasing from left to right.
As the value of $\sigma$ increases at fixed value of $\ln \mu$, the log-normal distribution is more and more skewed.
The sampling is ordering-agnostic only if $\sigma\ll |\ln \mu|$, i.e.\ for quasi-degenerate neutrinos.
For $\sigma \gtrsim |\ln \mu|$, when hierarchical spectra become allowed, there is a clear preference for the variable $(L^2-M^2)-(M^2-S^2)$ to take positive values, and hence there is an {\it a priori} preference for normal ordering.
The preference is quantified as the ratio of the occurences corresponding to NO and IO, $\mathrm{NO:IO}$, and reported in the top-right corner of each panel in Fig.~\ref{fig:hist} for that specific instance. We also provide an estimate of the associated error by repeating the analysis for $n=100$ times. We get $\mathrm{NO:IO}=(1.00\pm0.02)$, $(1.28\pm0.03)$, $(3.14\pm0.07)$, $(16.1\pm0.8)$ in the case of $\sigma=6\times 10^{-4},\,0.13,\,0.77,\,4.6$, respectively.
Note that the results do not depend on the specific choice of $\ln \mu$, i.e., we get the same ratio in favour of NO when considering different values of $\ln \mu$.

We checked that a symmetric distribution about $(L^2-M^2)-(M^2-S^2)=0$ (and $\mathrm{NO:IO=1}$) is obtained in case of a \emph{normal} prior distribution~%
\footnote{In this case, care must be taken so that the support of the prior distribution is positive, i.e., $X>0$ with $X=L,\,M,\,S$. This is obtained automatically provided that $\sigma\ll\mu$.
},
in contrast with the \emph{log-normal} one.
If $X=L,\,M,\,S$ are drawn from the same uniform distribution over the mass (case \textbf{B} in Sec.~\ref{subs:param}) or over the logarithm of the mass (case \textbf{C} in Sec.~\ref{subs:param}), we still note a preference for the normal ordering.
This preference is quantified as $\mathrm{NO:IO=1.6}$ when the mass prior is taken to be uniform, and $\mathrm{NO:IO=2.7}$ when the mass prior is taken to be uniform in the logarithm of the mass.
Such a mild preference can be explained by the same argument that we use to justify the preference for NO in the case of $\sigma \gtrsim \mu$ above: the uniform prior, either over the mass or the logarithm of the mass, allows hierarchical spectra to be drawn.

In the case of ii), the ordering is assigned depending on the relative magnitude of two numbers drawn from the same distribution (case \textbf{D} and \textbf{E} in Sec.~\ref{subs:param}).
One could check that the difference of two variables extracted from the same distribution, is itself a random variable with symmetric distribution about the origin.
This is true regardless of whether the splittings are drawn from a uniform, Gaussian or log-normal prior distribution.
As a result, in the case of ii), there is no \emph{a priori} preference for any of the two mass orderings.

The discussion in this section can help to better understand the pattern in the Bayes factor across different parameterizations visible in Figs.~\ref{fig:Bayes_factors}-\ref{fig:Bayes_factors_dchi2}.
In fact, cases \textbf{A}, \textbf{B} and \textbf{C} (representative of the class i) above) show a higher preference overall for NO with respect to cases \textbf{D} and \textbf{E} (representative of the ordering-agnostic class ii) above).
Within the former group (\textbf{A-C}), cases \textbf{A} and \textbf{C} show a larger Bayes factor than case \textbf{B} if we compare points obtained with the same dataset.

\section{Final remarks}
\label{sec:conclusions}

In this paper, we provide a picture of the state-of-the-art knowledge of the neutrino mass ordering.
We quantify the preference in favour of the normal ordering (NO), while accounting for data from neutrino oscillation experiments, the beta-decay experiment KATRIN, and from cosmological probes.
We thoroughly discuss the impact that different parameterization and prior choices might have on the results.

The neutrino mass ordering is one of the fundamental unknowns in the lepton sector of the standard model.
Its precise knowledge will have an enormous impact in our current understanding of particle masses, and in the underlying symmetries between quarks and leptons, both key ingredients towards a complete theory of particles and fields.
Another key ingredient that also plays a major role in our description of nature is the putative Majorana neutrino character.
Plenty of effort in the experimental neutrino community is being devoted to the design of future neutrinoless double beta decay experiments, which aim to extract the neutrino character.
However, the scale of these future projects strongly relies on the neutrino mass ordering: while next generation neutrinoless double beta decay experiments will likely be able to test the inverted ordering region, major improvements should be envisaged to fully cover the region allowed in normal ordering.

Given the importance that the neutrino mass ordering has in the neutrino scientific community (at the experimental, theoretical, phenomenological and strategical levels),
plenty of caution is needed before strong claims about the present \emph{data} preference for a given mass ordering are given.
Based on the analyses performed in this work, our main messages
are
(\emph{i}) there is a large difference in the Bayes factors, depending on the strategy followed to sample the neutrino mass parameter space, which could shift the preference for normal mass ordering even from a weak to a strong level;
(\emph{ii})
we therefore encourage the use of priors and parameterizations that are ordering-agnostic,
i.e.\ do not favor any mass ordering \emph{a priori}, and/or are less informative, i.e.\ maximize the information gain from the prior to the posterior.
This is the case for sampling over two mass splittings plus a mass scale, either via uniform or log-uniform distributions, or for sampling over the three individual neutrino masses \emph{via uniform prior distributions only};
(\emph{iii}) the current most robust result (i.e.\ driven exclusively by the data) is that the significance in favor of the normal mass ordering is $2.7\sigma$, from the combination of oscillations and cosmological data, which is conventionally not considered as ``conclusive evidence'',
and, finally,
(\emph{iv}) neutrino oscillation measurements are dominating the current moderate preference for normal mass ordering, while the most stringent cosmological measurements only strengthen the significance by $0.6\sigma$.
A weakened bound on the neutrino mass sum from cosmology, such as those obtained in extended models, would translate to reduced significance, in line with the oscillation data alone.

The above items are intended to clarify and explain
the status of current strong claims favouring the normal neutrino mass ordering.

\begin{acknowledgments}
We thank Alan Heavens and Sunny Vagnozzi for interesting discussions and comments.
SG acknowledges financial support from the European Union's Horizon 2020 research and innovation programme under the Marie Skłodowska-Curie grant agreement No 754496 (project FELLINI).
TB was supported through the INFN project GRANT73/Tec-Nu and from the COSMOS network (www.cosmosnet.it) through the Italian Space Agency (ASI) Grants 2016-24-H.0 and 2016-24-H.1-2018.
The work of OM is supported by the Spanish grants PID2020-113644GB-I00, PROMETEO/2019/083.
OM and TS acknowledge support from the European ITN project HIDDeN (H2020-MSCA-ITN-2019/860881-HIDDeN).
KF gratefully acknowledges support from the Jeff and Gail Kodosky Endowed Chair in Physics at the University of Texas, Austin; the U.S. Department of Energy, Office of Science, Office of High Energy Physics program under Award Number DE-SC-0022021 at the University of Texas, Austin; and the Vetenskapsrådet (Swedish Research Council) through contract number 638-2013-8993 at Stockholm University.
CAT is supported by the research grant ``The Dark Universe: A Synergic Multimessenger Approach'' number 2017X7X85K under the program ``PRIN 2017'' funded by the Ministero dell'Istruzione, Universit\`a e della Ricerca (MIUR).
MT acknowledges financial support from the Spanish grant PID2020-113775GB-I00 (AEI / 10.13039/501100011033).
\end{acknowledgments}

\bibliography{reply}

\end{document}